\documentstyle[sprocl]{article}

\def\be{\begin{equation}}
\def\ee{\end{equation}}
\def\bea{\begin{eqnarray}}
\def\eea{\end{eqnarray}}

\def\Lag{{\cal L}}
\def\op{{\cal O}}

\newcommand{\artref}[5]{{\rm #1}, {\rm #2} {\bf #3}, {\rm #4 (#5)}}

\begin{document}

\def\thefootnote{\fnsymbol{footnote}} 

\vspace*{-1.2cm}  
\begin{flushright}
CPT-2000/P.4072
\\
hep-ph/0010330
\\
October 2000 
\end{flushright}

\vspace*{0.4cm}
\title{CHIRAL DYNAMICS IN THE ELECTROWEAK SECTOR\footnote{Presented at
the 3rd Workshop on Chiral Dynamics -- Chiral Dynamics 2000: Theory
and Experiment, Newport News, VA, USA, 17-22 July 2000. To be
published in the Proceedings. }}

\author{A.\ NYFFELER}

\address{Centre de Physique Th\'{e}orique, CNRS-Luminy, Case 907 \\
    F-13288 Marseille Cedex 9, France \\ 
    E-mail: nyffeler@cpt.univ-mrs.fr
} 


\maketitle\abstracts{Using a chiral Lagrangian we show that strongly
interacting models of electroweak symmetry breaking are not in
conflict with precision data. Such models, like Technicolor, need not
lead to a heavy Higgs-like signal.  Furthermore, the allowed values
for the low-energy constants in the effective Lagrangian, derived from
bounds on the oblique correction parameters $S,T,U$, are not
unnatural. Finally, we point out that there are some problems with
gauge invariance, if one tries to relate the oblique parameters to the
low-energy constants in the ordinary chiral Lagrangian for QCD of
Gasser and Leutwyler. In particular, $S$ cannot be identified with
$l_5^{GL}$.}

\renewcommand{\thefootnote}{\arabic{footnote}}
\setcounter{footnote}{0}

\vspace*{-0.7cm}
\section{The electroweak chiral Lagrangian}

If the electroweak (EW) symmetry is dynamically broken by some
strongly interacting underlying theory, similarly to chiral symmetry
breaking in QCD, and if there is a mass gap between the observed
particles in the Standard Model (SM) and the scale of this underlying
theory, one can construct an effective field theory in analogy to
chiral perturbation theory. In the bosonic sector, the corresponding
EW chiral Lagrangian is of the form $\Lag_{eff} = \Lag_2 +
\Lag_4 + \ldots$, where $\Lag_k$ is of order $p^k$. The Lagrangian
$\Lag_2$ describes a gauged non-linear sigma model. $\Lag_4 =
\sum_i a_i \op_i$, where the $SU(2)_L \times U(1)_Y$ gauge-invariant
operators $\op_i$ are built from the light fields, i.e.\ the photon
and the $W$- and $Z$-bosons~\cite{EWchiralLag,EWChPT_gaugeinv}. The
low-energy constants (LEC's) $a_i$ parametrize different underlying
theories.

\vspace*{-0.2cm}
\section{Are strongly interacting models ruled out by precision data ?}
{\bf a) Light Higgs:} A heavy Higgs boson is excluded since all recent
SM fits of precision data point to a low Higgs mass, $M_H < 170~{\rm
GeV}$ at 95\% 
C.L. However, it is not true that all strongly
interacting theories lead to a heavy Higgs-like signal. As shown in
Table~\ref{tab:a_i_SM_TC}, the pattern of LEC's $a_i$ for a heavy
Higgs boson differs from the one in a simple Technicolor model,
estimated using VMD~\cite{LC_Note}.
\vspace*{-0.5cm}
\begin{table}[h]
\caption{Non-vanishing, renormalized LEC's $a_i^r(\mu=M_Z)$ for the SM
with $M_H = 1~{\rm TeV}$ and for a two-flavor, QCD-like Technicolor
(TC) model with a Technirho mass of $2~{\rm TeV}$.}
\begin{center}
\renewcommand{\arraystretch}{1.1}
\begin{tabular}{|l|r@{.}l|r@{.}l|r@{.}l|r@{.}l|r@{.}l|r@{.}l|}
\hline
 & \multicolumn{2}{|c|}{{$10^3 \times a_0$}}  
 & \multicolumn{2}{|c|}{{$10^3 \times a_1$}}  
 & \multicolumn{2}{|c|}{{$10^3 \times a_2$}}  
 & \multicolumn{2}{|c|}{{$10^3 \times a_3$}}  
 & \multicolumn{2}{|c|}{{$10^3 \times a_4$}}  
 & \multicolumn{2}{|c|}{{$10^3 \times a_5$}}  
\\ \hline  
SM &\hspace*{2mm}-11 & 8 &\hspace*{3mm}-2 & 6 &\hspace*{3mm}-0 & 8
&\hspace*{3.3mm}0 & 8 &\hspace*{3.3mm}1 & 6 &\hspace*{3.3mm}8 & 9 \\ 
TC & -14 & 7 & -8 & 9 & -5 & 4 & 5 & 4 & 5 & 2 & -0 & 3 \\  
\hline
\end{tabular}
\label{tab:a_i_SM_TC}  
\end{center}
\end{table}

\newpage
\noindent
{\bf b) $S$-parameter:} The oblique correction parameters $S,T,U$
describe effects of new physics {\em beyond\ } the SM that enter the
self-energies of EW gauge bosons~\cite{STU}. The PDG~\cite{PDG2000}
quotes the value $S = -0.16 \pm 0.11$ for $M_H = 300~{\rm GeV}$,
whereas estimates for Technicolor models lead to $S \approx {\cal O}
(1)$. The parameters $S,T,U$ can be related~\cite{Dobado_etal} to the
LEC's $a_i$, e.g., $S = 16 \pi \left[ - a_1(\mu) + \mbox{EWChL
loops$(\mu)$} \right].$ Interpreting the PDG values as bounds on the
deviations from the SM, $\Delta S \equiv S - S_{SM}(M_H)$, one obtains
with the values for the $a_i$ for a heavy Higgs boson:
\begin{displaymath}
\begin{array}{rclcl}
\Delta S & = & 16 \pi \left[ - a_1(\mu) -
{1\over 12} {(1/6) + \ln (M_H^2/\mu^2) \over 16 \pi^2} \right] &
\Rightarrow & 10^3 \times a_1(M_Z) = 1.8 \pm 2.2 \ , \\
\Delta T & = & {8 \pi \over c_W^2} \left[a_0(\mu) +
{3\over 8} {(1/6) + \ln (M_H^2/\mu^2) \over 16 \pi^2} \right] &
\Rightarrow & 10^3 \times a_0(M_Z) = -6.4 \pm 4.3 \ , \\
\Delta U & = & 16 \pi a_8  
& \Rightarrow & 10^3 \times a_8(M_Z) = 2.4 \pm 3.0 \ . 
\end{array} 
\end{displaymath}  
%
%
Although this rules out QCD-like Technicolor models, mainly from
$a_1$, the size of the $a_i$ is not unnatural for strongly
interacting theories, see also
Refs.~\cite{Dobado_etal,BaggerFalkSwartz}.

\section{$S \neq l_5^{GL}$ or the issue of gauge invariance}

There are some subtle problems with gauge
invariance~\cite{EWChPT_gaugeinv}, if one tries to relate $S$ to $a_1$
and $a_1$ to $l_5^{GL}$ in the ordinary chiral
Lagrangian~\cite{GL84}. There is a {\em qualitative} difference to
ChPT because the gauge fields in the EW chiral Lagrangian are
dynamical. Using the equations of motion for the gauge fields one can
in fact remove the operators corresponding to $a_1$ and $a_8$ from the
basis~\cite{EWChPT_gaugeinv}. Therefore, one cannot simply map
estimates for the LEC's $l_i^{GL}$ from QCD or from models into the
LEC's $a_i$ without performing a complete matching calculation.

\section*{Acknowledgments}
I would like to thank A.\ Schenk for the collaboration on some of the
topics presented here.  This work was supported by Schweizerischer
Nationalfonds.


\end{document}